\documentstyle[12pt]{article}
\newcommand{\be}{\begin{equation}}
\newcommand{\ee}{\end{equation}}
\newcommand{\bea}{\begin{eqnarray}}
\newcommand{\eea}{\end{eqnarray}}
\newcommand{\nn}{\nonumber}

\begin{document}

\def\gamh{\Gamma_H}
\def\esp #1{e^{\displaystyle{#1}}}
\def\de{\partial}
\def\eb{E_{\rm beam}}
\def\deb{\Delta E_{\rm beam}}
\def\sigm{\sigma_M}
\def\sigmmax{\sigma_M^{\rm max}}
\def\sigmmin{\sigma_M^{\rm min}}
\def\sige{\sigma_E}
\def\dsigm{\Delta\sigma_M}
\def\mh{M_H}
\def\lyear{L_{\rm year}}

\def\wstar{W^\star}
\def\zstar{Z^\star}
\def\ie{{\it i.e.}}
\def\etal{{\it et al.}}
\def\eg{{\it e.g.}}
\def\pzero{P^0}
\def\mt{m_t}
\def\mpzero{M_{\pzero}}
\def\mev{~{\rm MeV}}
\def\gev{~{\rm GeV}}
\def\gam{\gamma}
\def\lsim{\mathrel{\raise.3ex\hbox{$<$\kern-.75em\lower1ex\hbox{$\sim$}}}}
\def\gsim{\mathrel{\raise.3ex\hbox{$>$\kern-.75em\lower1ex\hbox{$\sim$}}}}
\def\ntc{N_{TC}}
\def\epem{e^+e^-}
\def\tauptaum{\tau^+\tau^-}
\def\lplm{\ell^+\ell^-}
\def\anti{\overline}
\def\mw{M_W}
\def\mz{M_Z}
\def\fbi{~{\rm fb}^{-1}}
\def\mupmum{\mu^+\mu^-}
\def\rts{\sqrt s}
\def\sigrts{\sigma_{\tiny\rts}^{}}
\def\sigrtssq{\sigma_{\tiny\rts}^2}
\def\sigrtsprime{\sigma_{E}}
\def\nsigrts{n_{\sigrts}}
\def\gampzero{\Gamma_{\pzero}}
\def\pzerop{P^{0\,\prime}}
\def\mpzerop{M_{\pzerop}}

\font\fortssbx=cmssbx10 scaled \magstep2
%
%
%
\hfill
%
%

%
\medskip
\begin{center}

{\Large\bf\boldmath Pseudoscalar Masses in the Effective  Theory for
Color-Flavor Locking in high density QCD \\}
\rm
\vskip1pc
{\Large R. Casalbuoni$^{a,b}$ and R. Gatto$^c$\\}

\vspace{5mm}
{\it{$^a$Dipartimento di Fisica, Universit\`a di Firenze, I-50125
Firenze, Italia
\\
$^b$I.N.F.N., Sezione di Firenze, I-50125 Firenze, Italia\\
$^c$D\'epart. de Physique Th\'eorique, Universit\'e de Gen\`eve,
CH-1211 Gen\`eve 4, Suisse}}
\end{center}
\bigskip
\begin{abstract}

\noindent

The constants
of the effective action
describing the massless modes in the color-flavor-locking
 phase of QCD  have been recently
 evaluated at high density. Values of the pseudoscalar masses for
nonvanishing quark masses have also been given. The calculated
values show however a puzzling feature: when $m_u=m_d=0$ and
$m_s\not=0$ all the goldstones continue to be massless. We show that an
additional invariant is present which avoids this feature. We
costruct this invariant and discuss the emerging mass pattern for
the pseudogoldstones. This provides for a complete scheme for the
pseudogoldstone masses in the color-flavor-locking  phase of QCD.

\end{abstract}
\newpage
\section{Introduction}

In a recent interesting paper \cite{son} the quantities appearing
in the effective action that we had proposed \cite{CG} for the
description of the massless modes in the color-flavor-locking
(CFL) phase of QCD
\cite{wilczek,wilczek1,alford,highdensity,literature}  have been
evaluated in the perturbative limit of very high density.
Furthermore an evaluation of the pseudoscalar masses generated
from nonvanishing quark masses in the QCD lagrangian has been
given.

First let us discuss the expected scaling of the pseudoscalar masses
 with  quark masses in the new phase. We know the standard
arguments at zero density giving the square mass of the
pseudoscalar as proportional to to quark mass times the
condensate $\langle\bar\psi\psi\rangle$. However, in the CFL
phase, for massless quarks, there is a discrete symmetry $Z_{2L}$
(a change of sign for the left-handed quarks) which requires the
condensate to be zero \cite{wilczek}. When we add the symmetry
breaking induced by the quark masses the previous statement is no
longer valid. Nevertheless we can still think of the condensate
 vanishing with the quark masses.

This
leads to the idea that higher dimension operators should be
considered, leading to a quadratic dependence of the pseudogoldstone masses
on the quark masses. However, also the mechanism discussed above,
that is the usual Adler-Dashen theory, when implemented with the idea
of a condensate vanishing for massless quarks, can give the
quadratic dependence.

In fact, in ref. \cite{son} it has been
shown that starting from the mass term in the QCD lagrangian
\be
{\cal L}_{m}=\psi_L^\dagger M\psi_R+{\rm h.c.}
\ee
and considering a mass matrix proportional to the identity matrix
$M=m 1_3$, one gets at very high density a shift of the vacuum energy
given by
\be
\Delta {\cal E}_{\rm vac}=- \frac{9\mu^2}{8\pi} m^2+{\rm h.c.}
\ee
with $\mu$ the chemical potential (the calculation is made
perturbatively since we are considering the limit of very  high
density). This shows that for such a mass matrix the condensate
is given by
\be
\sum_{u,d,s} \langle\psi_R^\dagger\psi_L\rangle=-\frac{9\mu^2}
{8\pi} m
\ee
Therefore, at the level of the effective action of the pseudogoldstones,
one can expect that there are different contributions to the masses.
One contribution from the usual Adler-Dashen mechanism, plus others
coming from higher dimension operators. One  higher dimension operator
is
 explicitly constructed in ref. \cite{son}. An important role is played
by  the $U(1)_A$ symmetry. Such a symmetry is  effectively
restored \cite{rapp2} at large values of $\mu$.

In this note we will show that
there is another term in the effective
lagrangian, which gives rise to a pseudoscalar mass pattern of the usual
hierarchical type,  and it still
respects all the needed symmetries.

Most importantly, this new term
solves a problem present in the paper \cite{son}. In fact the
operator considered in \cite{son} turns out to give that all
pseudoscalar masses continue to vanish for
$m_u=m_d=0$ even when $m_s\not=0$. This  result would be very puzzling.
A symmetry breaking term ($m_s\not=0$) in the fundamental lagrangian
would have
no effect in the effective action.  We cannot expect that, when an explicit
strange quark mass is introduced, all the massless goldstones,
of the situation where all quarks were massless, do remain massless.
We must expect
that those
which are no longer goldstones aquire a mass.  As we
shall see the additional invariant that we will construct
gives pseudoscalar masses which all vanish only when the strange quark mass
 mass vanishes. This avoids the puzzle.

\section{The effective theory}
We shortly review the effective action for the color-flavor
locked (CFL) phase of QCD introduced in \cite{CG}. The symmetry
breaking pattern is \cite{wilczek,wilczek1,alford}
\be
G=SU(3)_c\otimes SU(3)_L\otimes SU(3)_R \otimes U(1)\to
H=SU(3)_{c+L+R}
\ee
from dynamical  condensates
\be
\langle\psi_{ai}^L\psi_{bj}^L\rangle=-\langle\psi_{ai}^R\psi_{bj}^R
\rangle=\gamma_1\delta_{ai}\delta_{bj}+\gamma_2\delta_{aj}\delta_{bi}
\label{2}
\ee
($\psi_{ai}^{L(R)}$ are Weyl  spinors and  spinor indices are
summed). The indices $a,b$ and $i,j$ refer to $SU(3)_c$ and to
$SU(3)_L$ (or $SU(3)_R$) respectively. To be precise
 $H$ contains
an additional
$Z_2$, which  plays an essential role.

The effective lagrangian describes the CFL phase for
momenta smaller than the energy gap (existing
estimates range between 10 and 100 $MeV$).  Before  gauging
 $SU(3)_c$  we need 17 goldstones. We define coset matrix fields
$X$ ($Y$) transforming under $G$  as a left-handed (right-handed)
quark. We require
\be
X\to g_c X g_L^T,~~~~~Y\to g_c Y g_R^T
\ee
with $g_c\in SU(3)_c,~~~g_L\in SU(3)_L,~~~g_R\in SU(3)_R$.
$X$ and $Y$ are $SU(3)$ matrices, breaking respectively
$SU(3)_c\otimes SU(3)_L$ and $SU(3)_c\otimes SU(3)_R$.
 An additional goldstone is related to the
breaking of  baryon number. It corresponds to a $U(1)$
factor transforming under $G$ as
\be
U\to g_{U(1)} U,~~~~g_{U(1)}\in U(1)
\ee
We define the anti-hermitian and traceless
currents
\be
J_X^\mu=X\de^\mu X^\dagger,~~~ J_Y^\mu=Y\de^\mu Y^\dagger,~~~
J_\phi=U\de^\mu U^\dagger
\ee
which transform under
 $G$ as
\be
J_X^\mu\to g_c J_X^\mu g_c^\dagger,~~~J_Y^\mu\to g_c J_Y^\mu
g_c^\dagger,~~~J_\phi^\mu\to J_\phi^\mu
\ee
At finite
density Lorentz  invariance is broken.
 Barring  WZW terms  (see ref. \cite{CG} for a brief
discussion), the most general $O(3)$ symmetric  lagrangian
invariant under $G$, with at most two derivatives, is
\bea
{\cal L}&=&-\frac{F_T^2}4 Tr[(J_X^0-J_Y^0)^2]-
\alpha_T\frac{F_T^2}4 Tr[(J_X^0+J_Y^0)^2] -\frac{f_T^2}2 (J_\phi^0)^2\nn\\
&+&\frac{F_S^2}4 Tr[(\vec J_X-\vec J_Y)^2]+
\alpha_S\frac{F_S^2}4 Tr[(\vec J_X+\vec J_Y)^2] +\frac{f_S^2}2 (\vec J_\phi)^2
\label{BESS}
\eea
We have required  invariance under parity (symmetry under
$X\leftrightarrow Y$).

With the parametrization
\be
X=\esp{i\tilde\Pi_X^aT_a},~~~
Y=\esp{i\tilde\Pi_Y^aT_a},~~~ U=\esp{i\tilde\phi},~~~
a=1,\cdots 8
\ee
(the  $SU(3)$ matrices $T_a$  satisfy
$Tr[T_aT_b]=\frac 1 2 \delta_{ab}$)
and by defining
\be
\Pi_X=\sqrt{\alpha_T}\,\frac{F_T}2(\tilde\Pi_X+\tilde\Pi_Y),~~~
\Pi_Y=\frac{F_T} 2(\tilde\Pi_X-\tilde\Pi_Y),~~~
\phi=f_T\tilde\phi
\label{rescaling}
\ee
 the  kinetic term is
\be
{\cal L}_{\rm kin}=\frac 12 ({\dot\Pi_X}^a)^2+
\frac 12
({\dot\Pi_Y}^{a})^2+ \frac 1 2 (\dot\phi)^2
-\frac {v_X^2}2 |\vec\nabla\Pi_X^{a}|^2-
\frac {v_Y^2}2
|\vec\nabla\Pi_Y^{a}|^2- \frac {v_\phi^2} 2 |\vec\nabla\phi|^2
\ee
where
\be
v_X^2=\frac{\alpha_S F_S^2}{\alpha_T F_T^2},~~~v_Y^2=\frac{F_S^2}
{F_T^2},
~~~v_\phi^2=\frac{f_S^2}{f_T^2}
\label{velocity}
\ee
The three types of goldstones satisfy linear
dispersion relations $E=vp$,
with different velocities,

For local $SU(3)_c$
invariance we  make the derivatives covariant
\be
\de_\mu X\to D_\mu X= \de_\mu X-g_\mu X,~~~
\de_\mu Y\to D_\mu Y= \de_\mu Y-g_\mu Y,~~~ g_\mu\in {\rm Lie}~SU(3)_c
\ee
The  currents become
\be
J_X^\mu=X\de^\mu X^\dagger+g^\mu,~~~ J_Y^\mu=Y\de^\mu
Y^\dagger+g^\mu
\ee
giving for the lagrangian
\bea
{\cal L}&=&-\frac{F_T^2} 4Tr[(X\de^0 X^\dagger -Y\de^0
Y^\dagger)^2]-\alpha_T\frac{F_T^2} 4Tr[(X\de^0 X^\dagger +Y\de^0
Y^\dagger+2g^0)^2]\nn\\ &-&\frac{f_T^2}2(J_\phi^0)^2+{\rm
spatial~
 terms~and~kinetic~part~for}~ g^\mu
\label{lagr BESS}
\eea
We introduce
\be
g_\mu=ig_s\frac{T_a}2 g_\mu^a
\ee
where $g_s$  is the QCD coupling constant. The gluon
field becomes massive. This can  be easily seen in the unitary gauge
$X=Y^\dagger$,
where
\be
\tilde\Pi_X=-\tilde\Pi_Y
\ee
or
\be
\Pi_X=0,~~~\Pi_Y=F_T\tilde\Pi_X
\ee
 The gluon mass (for the expected velocities of
order one) is
\be
m_g^2=\alpha_T g_s^2\frac {F_T^2}4
\ee
The $X\leftrightarrow Y$ symmetry, in this
gauge, implies $\Pi_Y\leftrightarrow -\Pi_Y$.

The gluon kinetic term can be neglected for energies much smaller
than the gluon mass. The lagrangian (\ref{lagr BESS})
is then the hidden gauge symmetry version of the chiral QCD
lagrangian   (except for the  field
$\phi$). In fact, in this limit, the gluon field becomes
auxiliary  and  can be eliminated through its equation of
motion
\be
g_\mu=-\frac 1 2(X\de_\mu X^\dagger+Y\de_\mu Y^\dagger)
\ee
obtaining
\be
{\cal L}=-\frac{F_T^2}4Tr[(X\de^0 X^\dagger-Y\de^0
Y^\dagger)^2]-\frac{f_T^2} 2(J_\phi^0)^2+ {\rm spatial~terms}
\label{QCD}
\ee
or
\be
{\cal L}=\frac{F_T^2}4 \left(Tr[\dot \Sigma\dot \Sigma^\dagger]-
v_Y^2 tr[\vec\nabla \Sigma\cdot\vec\nabla \Sigma^\dagger]\right)
-\frac{f_T^2}2\left((J_\phi^0)^2-v_\phi^2|\vec J_\phi|^2\right)
\ee
where $\Sigma=Y^\dagger X$ transforms under the group $SU(3)_c\otimes
SU(3)_L\otimes
SU(3)_R$ as $\Sigma\to g_R^* \Sigma g_L^T$. The goldstone $\phi$
 could be interpreted as a particularly light dibaryon state,
 $(udsuds)$, considered by R. Jaffe \cite{jaffe}.  After the breaking of
color, one has
 the massless photon and 9 physical goldstones
transforming as  $1+8$ under the unbroken SU(3).

In our previous paper \cite{CG} we did not consider the pseudogoldstone mode
associated to the $U(1)_A$ symmetry since it is not massless.
However, in the present context it plays an important role. Its
kinetic term can be introduced in complete analogy with what we
have done for the mode associated to the vector $U(1)$. The
contribution to the lagrangian is given by
\be
-\frac{f_A^2} 2((J_\theta^0)^2-v_\theta^2|\vec J_\theta|^2)
\ee
where
\be
J_\theta^\mu=V\de^\mu V^\dagger, V=\esp{i\tilde\theta}
\ee
and, for the correct normalization of the kinetic term, one has
to define the field
\be
\theta=\frac{1}{f_A}\tilde\theta
\ee
Under a $U(1)_A$ transformation all the fields are invariant
except for
\be
V\to \esp{i\alpha}V, ~~~\esp{i\alpha}\in U(1)_A
\label{axial}
\ee
An equivalent description can be obtained by including the $U(1)$
fields into  $X$ and  $Y$ fields belonging to $U(3)$, but we will
stay with the  formalism just described.

\section{Mass terms}
We are now in the position of discussing the mass terms for the
pseudogoldstones.

The quark mass term in the QCD lagrangian is given by
\be
{\cal L}_{m}=\psi_L^\dagger M\psi_R+{\rm h.c.}
\ee
By thinking of $M$ as a set of external fields, the QCD
lagrangian preserves all its symmetries if we require that under
the global group $SU(3)_L\otimes SU(3)_R\otimes U(1)\otimes
U(1)_A$ the fields $M$ transform as
\be
M\to \esp{2i\alpha}g_L Mg_R^\dagger
\ee
As said before, we insist on keeping the $U(1)_A$ \cite{son}
(in the CFL phase this symmetry is restored at very high
density). The effective lagrangian must respect this symmetry
extended to the external fields.  In ref. \cite{son} it has
been observed that an
invariant term is given by
\be
\Delta_1=-c\cdot det(M)\cdot Tr(M^{-1T}\Sigma)V^{-4}+{\rm h.c.}
\ee
(this term differs in form from the one of ref. \cite{son}
because, as explained above, our $\Sigma$ field is invariant
under $U(1)_A$).

The main observation of the present paper is to notice that at the order $M^2$
another invariant term exists. It reproduces the mass mechanism
of the usual low density chirally broken phase. In the present
case however the condensate has a dependence on the quark masses.
In fact, we notice that the quantity
\be
\Delta_2=-f(Tr(M^\dagger M))Tr[M^*\Sigma] V^2+{\rm h.c.}
\ee
is invariant, for any choice of the function $f$, since the
transformation of $V$ (see eq. (\ref{axial})) under $U(1)_A$
compensates the change in $M$. For an evaluation at high density
of the constant $c$ and of the function $f$ we make use of the
results of ref. \cite{son} for the region of very high densities.
We can then
 match the shift of the vacuum energy
in QCD to that of the effective theory. For a mass matrix
proportional to the unit matrix $M=m 1_3$, and for $M={\rm diag}
(0,0,m_s)$, one has in QCD, respectively
\be
\Delta{\cal E}_{\rm vac}=-\frac {9\mu^2}{8\pi} m^2,~~~ \Delta{\cal E}_{\rm vac}
=-\frac{3\mu^2}{8\pi} m_s^2
\ee
where $\mu$  is the chemical potential. Since $\Delta_1$ vanishes
for $m_d=m_u=0$, it follows  that
\be
f(Tr(M^\dagger M))=-\frac{3\mu^2}{8\pi}\sqrt{Tr(M^\dagger M)}
\ee
Then, choosing the case $M=m 1_3$, we get
\be
c=\frac{9(1-\sqrt{3})}{8\pi}\mu^2
\ee
The invariant $\Delta_2$ in the effective lagrangian solves the
puzzle discussed in the introduction. The pseudogoldstone masses
do no longer vanish for $m_u=m_d=0$ and for a massive strange
quark. Notice that the invariant $\Delta_2$ gives a mass
pattern identical to the usual one of zero density, except for
the factor arising from the function $f$, that is
$\sqrt{m_u^2+m_d^2+m_s^2}$. This  is  what one would expect from
an  Adler-Dashen type mechanism, except that in the present case
one has to think of a
 condensate which vanishes with the quark masses due to the $Z_2$
invariance, as iscussed in the Introduction. The invariant
$\Delta_2$ gives also a contribution to the mass of the
pseudogoldstone $\theta$ associated to the axial symmetry and to
its mixings with $\pi^0$ and $\eta$. On the other hand it does
not contribute to the mass of the pseudogoldstone $\phi$
associated to the barionic number. We notice that in the
phenomenological situation, where $m_u,m_d\ll m_s$, the invariant
$\Delta_2$ dominates the pseudogoldstone masses.

Using the results of ref. \cite{son} one finds that the pseudogoldstone
masses are completely determined by the quark masses. In fact, in
our notations, from  ref. \cite{son} we get
\be
F_T^2=(21-8\ln 2)\frac{\mu^2}{9\pi^2},~~~
f_T^2=\frac{9\mu^2}{\pi^2}\approx f_A^2
\ee
Since the pseudogoldstone masses scale as the inverse of the square of the
decay constants, one finds that they do not depend on the
chemical potential.

As a last observation we notice that for $m_u=m_d=0$ and
$m_s\not=0$, the mass pattern is the same as in the chiral phase
(except for a multiplicative constant). Therefore the pions
$\pi^\pm$ and $\pi^0$ remain massless, whereas all the other
pseudogoldstones become massive (excluding the $\phi$ particle).
Then, as expected, there are 3 massless Goldstone bosons coming
from the breaking of
\be
SU(3)_c\otimes SU(2)_L\otimes SU(2)_R\to SU(2)
\ee
(notice that 8 goldstones are eaten up by the gluons). Since it
has been shown that for two flavors at high density the symmetry
breaking pattern differs from the one in eq. (37) \cite{wilczek2},
we see that some interesting phenomenon has to happen in the
transition between the case of two and three flavors.

\section{Conclusions}

In this note we have discussed the pattern of the masses of the
pseudogoldstones of the color-flavor-locked phase of QCD at high
density when  quark masses are considered. We have examined the
different mass terms that have to be introduced in the effective
lagrangian. A previous puzzling feature, of all goldstones
remaining massless for $m_u=m_d=0$ and $m_s\not=0$ was due to
neglection of a mass invariant, whose structure reminds, in a
properly modified way, of the Adler-Dashen mechanism of the
chirally broken phase at low densities. The pseudogoldstone mass
pattern is then completely understood.
\begin{center}
{\bf Acknowledgements}
\end{center}
\medskip
We would like to thank Krishna Rajagopal for an enlightening
exchange of correspondence on the subject.

\end{document}